\documentclass[%
 reprint,
superscriptaddress,
 amsmath,amssymb,
 aps,
 prl,
]{revtex4-2}

\usepackage{graphicx}
\usepackage{dcolumn}
\usepackage{bm}
\usepackage{physics}
\usepackage{booktabs}
\usepackage{enumerate}

\usepackage[pdftex,pagebackref=false]{hyperref} 
\hypersetup{
    colorlinks=true,        
    linkcolor=blue,         
    citecolor=red,        
    filecolor=magenta,      
    urlcolor=cyan           
}

\usepackage{xcolor}
\usepackage{cleveref}
\usepackage{orcidlink}

\newcommand{\kket}[1]{\left|\left. #1 \right\rangle\right\rangle}
\newcommand{\bbra}[1]{\left\langle\left\langle #1 \right.\right|}
\newcommand{\I}{\mathrm i}

\DeclareMathOperator*{\argmin}{argmin} 
\begin{document}

\title{Circuit compression for $\mathbf2$D quantum dynamics}

\author{Matteo D'Anna\orcidlink{0000-0002-9426-0377}}\thanks{madanna@ethz.ch}
\affiliation{Institut f\"{u}r Theoretische Physik, ETH Z\"{u}rich,
Wolfgang-Pauli-Str. 27, 8093 Z\"{u}rich, Switzerland}
\author{Yuxuan Zhang\orcidlink{0000-0001-5477-8924}}
    \thanks{quantum.zhang@utoronto.ca}
    \affiliation{Department of Physics and Centre for Quantum Information and Quantum Control, University of Toronto,
60 Saint George St., Toronto, Ontario M5S 1A7, Canada}
    \affiliation{Vector Institute, W1140-108 College Street, Schwartz Reisman Innovation Campus
Toronto, \\Ontario
M5G 0C6, Canada}
\author{Roeland Wiersema\orcidlink{0000-0002-0839-4265}}\thanks{rwiersema@flatironinstitute.org}
\affiliation{Center for Computational Quantum Physics, Flatiron Institute, 162 Fifth Avenue, New York, NY 10010, USA}
\author{Manuel S. Rudolph\orcidlink{0000-0003-1261-1442}}
\thanks{manuel.rudolph@epfl.ch}
\affiliation{Institute of Physics, Ecole Polytechnique Fédérale de Lausanne (EPFL), Lausanne, Switzerland}
\affiliation{Centre for Quantum Science and Engineering, Ecole Polytechnique Fédérale de Lausanne (EPFL), Lausanne, Switzerland}
\author{Juan  Carrasquilla\orcidlink{0000-0001-7263-3462}}\thanks{jcarrasquill@ethz.ch}
\affiliation{Institut f\"{u}r Theoretische Physik, ETH Z\"{u}rich,
Wolfgang-Pauli-Str. 27, 8093 Z\"{u}rich, Switzerland}

\date{\today}

\begin{abstract}
\end{abstract}


\begin{abstract}
The study of out-of-equilibrium quantum many-body dynamics remains one of the most exciting research frontiers of physics, standing at the crossroads of our understanding of complex quantum phenomena and the realization of quantum advantage. Quantum algorithms for the dynamics of quantum systems typically require deep quantum circuits whose accuracy is compromised by noise and imperfections in near-term hardware. Thus, reducing the depth of such quantum circuits to shallower ones while retaining high accuracy is critical for quantum simulation. Variational quantum compilation methods offer a promising path forward, yet 
a core difficulty persists: ensuring that a variational ansatz $V$ faithfully approximates a target unitary $U$.
Here we leverage Pauli propagation techniques to develop a strategy for compressing circuits that implement the dynamics of large two-dimensional ($2$D) quantum systems and beyond. As a concrete demonstration, we compress the dynamics of systems up to $30 \times 30$ qubits and achieve up to $>1000$ times improvement in accuracies that surpass standard Trotterization methods 
at identical circuit depths. To experimentally validate our approach, we execute the compressed circuits on Quantinuum’s H1 quantum processor; we observe that, even with device noise, the measured physical observables are an order of magnitude more accurate than the ones obtained from Trotterized circuits under the same amount of quantum gate consumption.
Our high-dimensional circuit compression scheme brings us one step closer to a practical quantum advantage by allowing longer simulation times at reduced quantum resources and unlocks the exploration of large families of hardware-friendly ans\"atze.
\end{abstract}

\maketitle

As quantum devices approach the threshold for demonstrating quantum advantage~\cite{aaronson2011computational,preskill2012quantum,aaronson2013bosonsampling,childs2013universal,aaronson2016complexity,bouland2018quantum,movassagh2018efficient,aaronson2024verifiable,arute2019quantum,wu2021strong,king2025beyond,zhang2025complexity}, they continue to face limitations stemming from various noise sources, particularly those arising from entangling gates~\cite{preskill2018quantum,evered2023high,li2023error,zhang2023quantum,sete2024error}. Consequently, a critical question arises: how can one optimally utilize classical computational methods to reduce resource demands when executing quantum operations—a problem known as quantum circuit compilation~\cite{trotter1959product,suzuki1976generalized,barenco1995elementary,shende2005synthesis,cincio2018learning,khatri2019quantum,bagherimehrab2024faster}. 

Quantum dynamical simulation is widely recognized as a particularly promising candidate for demonstrating practical quantum advantage~\cite{hild2014far,zhang2017observation,scholl2021quantum,daley2022practical,kim2023evidence,wienand2024emergence,zhang2024observation,will2025probing,haghshenas2025digitalquantummagnetismfrontier,king2025beyond}. 
Here, deterministic compilation methods such as Trotterization have repeatedly been demonstrated to be suboptimal compared to fine-tuned variational quantum compilation (VQC) methods ~\cite{jones2022robust,cincio2018learning,khatri2019quantum,khatri2019quantum,cirstoiu2020variational,lin2021real,yao2021adaptiveVQD, kokcu2022algebraic,kokcu2022fixed,mc2023classically,tepaske2023optimal,mc2024towards,tepaske2024optimal,zhang2024scalable,zhang2025classical,mizuta2022local,kanasugi2023computation,kanasugi2024subspace,le2025riemannian, gibbs2025learning,gibbs2024deep}. Despite this potential of VQC, many recent quantum simulation results~\cite{kim2023evidence,haghshenas2025digitalquantummagnetismfrontier} still rely heavily on traditional Trotterization methods~\cite{trotter1959product,suzuki1976generalized}. The impracticality of VQC methods arises mainly from the scalability challenges faced by existing approaches. 

At the heart of this scalability issue is the difficulty in explicitly representing unitaries classically, especially in two-dimensional ($2$D) systems and beyond. This is particularly limiting because higher-dimensional systems represent precisely the regime in which quantum computing is expected to outperform classical simulation methods such as Time-Evolving Block Decimation (TEBD)~\cite{verstraete2004matrix,vidal2004efficient} in the near future. Recent progress, however, has provided a promising path forward: a series of quantum machine learning (QML) studies~\cite{Banchi2021generalization, huang2021power,caro2022generalization, caro2023learning,jerbi2023power} have shown that short-time quantum evolutions are highly amenable to being learned directly from data, providing a sample-efficient solution to the compilation challenge. There, the Hilbert-Schmidt inner product, $|\text{tr}(U^\dagger V(\vec\theta)) |^2/2^n$ where $U$ is a target unitary and $V(\vec\theta)$ is the VQC circuit, is approximated by a cost function based on sampling their action on a handful of random product states, $\mathbb{E}[ |\langle \psi | U^\dagger V(\vec\theta) | \psi \rangle|^2 ]$. 

Further, in Ref.~\cite{zhang2024scalable}, the authors leveraged this insight, introducing a scalable VQC framework by combining tensor-network techniques with the QML cost function.
They showed that, for shallow $U$, this approach enabled rapid classical simulation via matrix product states (MPS) for 1D and quasi-1D systems, achieving significant compression—up to a factor of 10 reduction in required gate resources compared to Trotterization methods. It has been shown that one could then repeatedly apply these compressed circuits to reliably generate long-time evolutions, just as in Trotterization~\cite{gibbs2024dynamical,zhang2024scalable}. 

In this work, we take a substantial leap forward by extending these machine-learning-inspired VQC techniques to large two-dimensional lattice systems. Specifically, we integrate this machine learning approach with Pauli propagation (PP) and employ a locality-based formalism in the cost function that perfectly matches the sparsity required by PP. Moreover, we develop a PP truncation scheme that is particularly suitable for simulating dynamics. We benchmark our algorithm across various lattice geometries, including heavy-hex~\cite{hetenyi2024creating} and conventional $2$D square lattices, with dimensions as large as $ 30 \times 30 $ and report consistent observation of orders-of-magnitude improvements in accuracy at comparable resource expenditures, which is especially pronounce when the Hamiltonian is time-dependent, as the Trotterization performs poorly. We show that our approach is a useful tool in dynamics simulations with current quantum hardware by implementing the dynamics of a cloud of hard-core bosons to high accuracy on the Quantinuum H1 chip and reaching accuracy far beyond Trotterization. 
\begin{figure*}
    \centering
    \includegraphics[width=.9\textwidth]{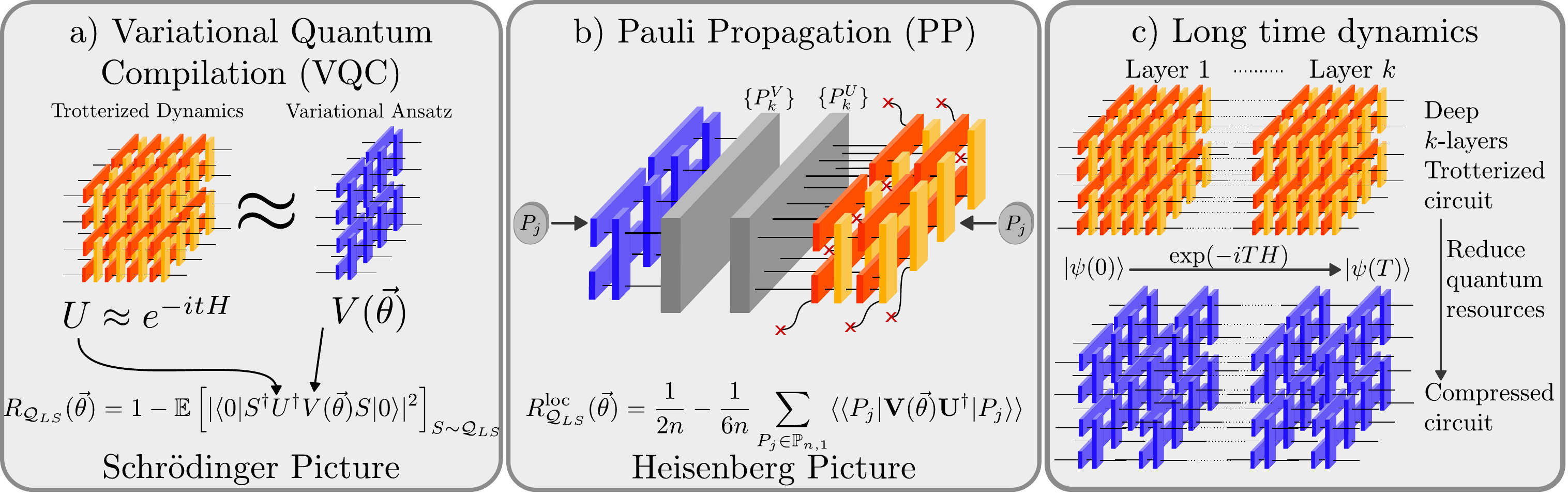}
    \caption{Overview of the VQC approach leveraged in this work. a) Starting in the Schr\"odinger picture, we approximate the Hilbert-Schmidt product between $U$ and $V$ by a global risk function obtained from sampling $U^\dagger V$'s action on product states. b) Shifting to the Heisenberg picture, we analytically show that the global risk function \eqref{eq:R_prod} can be reduced to an evaluation of expectation with local observables, $P_j$, as shown in \eqref{eq:locHaarpref}. In practice we compute $\bbra{P_j}\bm{V(\vec\theta)}$ and $\bm{U^\dagger}\kket{P_j}$ separately, and then take their inner product. c) The compressed circuit can then be applied $k$ times, accurately following the system's dynamics for long times $T$, and requiring much less quantum resources compared to a deep accurate Trotterization (bottom right).}
    \label{fig:overview}
\end{figure*}

\paragraph{Framework}\label{sec:setup}
In the context of quantum computing, the goal of variational compression algorithms is to take a target unitary $U$ acting on an $n$-qubit Hilbert space $(\mathbb{C}^2)^{\otimes n}$ and approximate it with a parametrized unitary $V(\vec\theta)$, where $\vec{\theta}\in\mathbb{R}^m$.
We target unitaries of the form $U\approx \exp(-i t H)$, generated by the Trotterization of the dynamics under some local Hamiltonian of interest $H$,
and show that lower-depth unitaries $V(\vec{\theta})$ can accurately approximate $U$ with significantly fewer resources. To determine the accuracy of the approximation, we start with the Hilbert-Schmidt inner product and define the cost
\begin{equation}\label{eq:hst}
    C_{\text{HST}}(\vec \theta) = 1- \frac 1{4^n}\left|\mathrm{tr}\left(U^\dagger V(\vec\theta)\right) \right|^2.
\end{equation}
In practice, minimizing the HST cost~\eqref{eq:hst} for unitaries acting on large systems is difficult, as explicitly writing down the unitary is generally hard due to the exponential scaling of the Hilbert space. However, it's possible to significantly simplify the optimization problem by considering a machine-learning-inspired approach.
We define the \textit{expected risk} on a set of random states that can be prepared by $S\ket{0}$, where $S$ is a random unitary drawn from some ensemble $\mathcal{Q}$ as
\begin{equation}\label{eq:RQ}
    R_{\mathcal{Q}}(\vec \theta)=\mathbb E \left[1-|\langle0 |S^\dagger U^\dagger  V(\vec \theta)S|0\rangle|^2 \right]_{S \sim \mathcal{Q}}.
\end{equation}
When $\mathcal{Q}$ is taken to be the global Haar ensemble, $\mathcal{Q}_{\text {Haar}_n}$, we have the relation~\cite{nielsen2002simple, horodecki1999teleport, khatri2019quantum}
\begin{equation}
    C_{\mathrm{HST}}(\vec \theta)=\frac{2^n+1}{2^n}R_{\mathcal{Q}_{\mathrm{Haar}_{\mathrm{n}}}}(\vec \theta).
\end{equation}
This cost is still challenging to evaluate, as global Haar random states have exponential circuit complexity on average. To circumvent this issue, we make use of the equivalence between ensembles~\cite{caro2023learning}

\begin{equation}\label{eq:odg}
    \frac 12 R_{\mathcal{Q}_{\mathrm{Haar}_n}}(\vec\theta) \leq \frac{2^n}{2^n+1} R_{\mathcal{Q}_{LS}}
    (\vec \theta) \leq R_{\mathcal{Q}_{\mathrm{Haar}_n}}(\vec\theta),
\end{equation}
where $\mathcal{Q}_{LS}$ is a locally scrambling ensemble~\cite{caro2023learning}. For example, we can take $\mathcal{Q}_{LS}$ to be the ensemble of products of \emph{single-qubit} Haar random unitaries.
By defining $L_S(\vec \theta):=S^\dagger U^\dagger  V(\vec \theta) S$, we rewrite the expected risk with respect to $\mathcal{Q}_{LS}$ (see \Cref{fig:overview}a) as
\begin{align}\label{eq:R_prod}
   R_{\mathcal{Q}_{LS}}(\vec \theta) = 1-\mathbb E\left[
   \mathrm{tr}(|0\rangle\langle 0|L_S(\vec \theta)|0\rangle \langle 0|L_S(\vec \theta)^\dagger)\right]_{S \sim \mathcal{Q}_{LS}}.
\end{align}

\paragraph{Analytical results}
In order to efficiently simulate the quantum dynamics of both the target $U$ and ansatz $V(\theta)$, we employ the numerical framework of Pauli Propagation (PP)~\cite{rall2019simulation, loizeau2025quantum, Begu_i__2025, fontana2025classical, rudolph2025pauli}. We now construct a local version of the expected risk in Equation~\eqref{eq:R_prod} using the Pauli transfer matrix (PTM) formalism. Here, a Hermitian operator $A$ is vectorized as $\kket{A}$ with components $\kket{A}_j = \tr(AP_j)/2^n$ where $P_j \in \mathbb{P}_n$ and $\mathbb{P}_n = \{I, X, Y, Z\}^n$. In this formalism, the inner product is defined as: $\bbra{A} \mathbf U \kket{B} :=  \tr\left(A^\dagger U B U^\dagger\right)/2^n$.

\Cref{eq:R_prod} in PTM notation becomes
\begin{equation}\label{eq:Rq_full}
    R_{\mathcal{Q}_{LS}}(\vec \theta) = 1 - \mathbb E\left[\bbra{0} \bm {L_S(\vec\theta)}\kket{0}\right]_{S \sim \mathcal{Q}_{LS}}.
\end{equation}
The all-zero state $\ketbra{0}=(I+Z)^{\otimes n}/2^n$ is a linear combination of exponentially many ($2^n$) Pauli strings, which prohibits efficient computation within the PTM framework.
Fortunately, Ref.~\cite{caro2023learning} shows that the replacement $\ketbra{0}\to \frac{1}{n}\sum_j (1+Z_j)/2$ leads to the \textit{local risk} 
\begin{equation}\label{eq:Rloc}
    R_{\mathcal{Q}_{LS}}^{\mathrm{loc}}(\vec \theta)= \frac{1}{2} - \frac{1}{2n} \sum_{j=1}^n\mathbb E\left[\bbra{0} \bm {L_S(\vec\theta)} \kket{Z_j}\right]_{S \sim \mathcal{Q}_{LS}}
\end{equation}
that satisfies the following equivalence between ensembles,
\begin{equation}
    \frac12R_{\mathcal{Q}_{LS}}^{\mathrm{loc}}(\vec \theta) \leq R_{\mathcal{P}_{LS}}(\vec \theta) \leq 2nR_{\mathcal{Q}_{LS}}^{\mathrm{loc}}(\vec \theta),
\end{equation}
where $\mathcal{Q}_{LS},\mathcal{P}_{LS}$ are locally scrambling ensembles.

The local risk \eqref{eq:Rloc} provides a reliable proxy for the true risk \eqref{eq:Rq_full} by tracking only a manageable number of Pauli strings---specifically, a number that grows linearly with the system size rather than exponentially which offers a substantial computational advantage. Additionally, we demonstrate that the averaging process over a locally scrambling distribution $\mathcal{Q}_{LS}$ can be avoided during simulation by replacing it with a compact analytical expression (derivation provided in the SM). This leads to the following simplified form illustrated in \Cref{fig:overview}b):
\begin{equation}\label{eq:locHaarpref}
    R_{\mathcal{Q}_{LS}}^{\mathrm{loc}}(\vec \theta)= \frac{1}{2} - \frac{1}{6n} \sum_{P_j\in\mathbb P_{n,1}} \langle\langle P_j \vert \bm{V(\vec\theta)U^\dagger}\vert P_j\rangle \rangle,
\end{equation}
where $\mathbb P_{n,w}:=\{P\in \mathbb P_n: W(P)=w\}$ is the subset Pauli strings with weight $w$.

In addition, if we consider compressing translation invariant (TI) systems with TI ans\"atze, \Cref{eq:locHaarpref} takes an even simpler form: for any two sites $j,k$ it holds $\bbra{0}\bm{L_S^\dagger}\kket{Z_j} = \bbra{0}\bm{L_S^\dagger}\kket{Z_k}$, and therefore \Cref{eq:locHaarpref} reduces to 
\begin{equation}\label{eq:ClocPPTI}
    R^\mathrm{loc,TI}_{\mathcal{Q}_{LS}}(\vec \theta) = \frac 12 -\frac1{6}\sum_{P=X_j,Y_j,Z_j} \langle\langle P  \vert \bm{V(\vec\theta)U^\dagger}\vert P\rangle \rangle.
\end{equation}
We have fully transformed the risk into a Heisenberg picture variant that can, in principle, be efficiently estimated using PP. While we could directly use PP to simulate $\bm{V(\vec\theta)U^\dagger}\vert P_j\rangle \rangle$, it is much more efficient to employ a ``meet in the middle'' approach: We compute $\langle\langle P_j \vert \bm{V(\vec\theta)}$ and $\bm{U^\dagger}\vert P_j\rangle \rangle$ separately, and then take their inner product.

\medskip
To summarize, we have reduced the problem of calculating the Hilbert-Schmidt inner product of~\Cref{eq:hst} to propagating $3n$ Pauli strings through our circuit. Compared to the previous QML-based cost functions~\cite{caro2022generalization,caro2023learning,zhang2024scalable}, ~\Cref{eq:locHaarpref} is exact and does not require sampling a finite number of ``data'' states, permitting much more efficient classical simulations. Putting everything together, we are left with the optimization problem
\begin{align}
\vec{\theta}^{\star}:=\argmin_{\vec{\theta}}\:R_{\mathcal{Q}_{LS}}^{\mathrm{loc}}(\vec \theta),
\end{align}
which we solve using a conjugate gradient descent algorithm using \texttt{OptimKit.jl}.
(see the Method section for a review on PP and truncation methods used to improve efficiency).
Once the parameters $\vec{\theta}^{\star}$ are determined, the ansatz $V(\vec{\theta}^{\star})$ can be executed $k$ times on a quantum device to reach the long-time dynamics at $T=k\cdot t$, significantly reducing the required quantum resources compared to a standard Trotterization approach (see \Cref{fig:overview}c).

\begin{figure*}
    \centering
    \includegraphics[width=.9\textwidth]{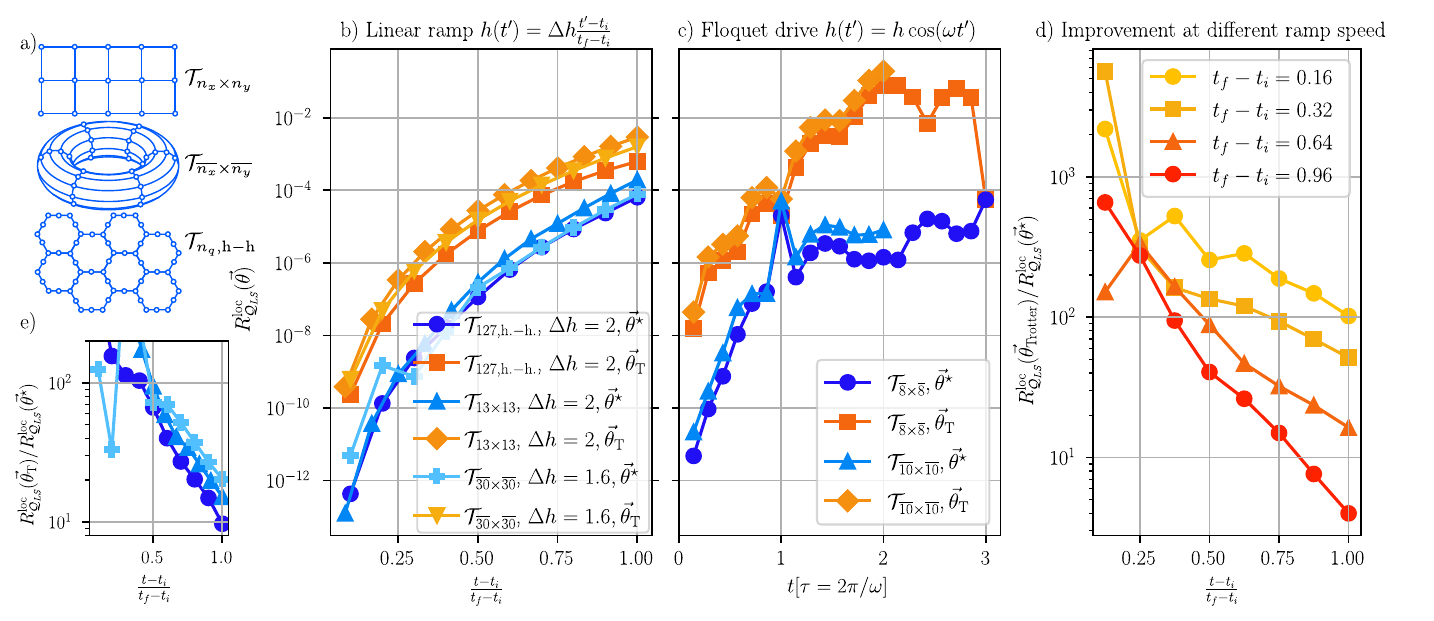}
    \caption{Compression for large 2D systems and comparison with Trotterization. At time $t$, the target $U$ cosists of $L_U\cdot t/\Delta t$ layers of Trotterization, whereas $V$ consists of $L_V$ layers at all times. a) Illustration of $\mathcal T_{n_x\times n_y}, \mathcal T_{\overline{n_x}\times \overline{n_y}}, \mathcal{T}_{n,\mathrm{h.-h.}}$. 
    In b)-d) we compare costs for the Trotter and compressed circuits. b) compression of the TFIM Hamiltonian, for $\mathcal{T}_{127,\mathrm{h.-h.}}$
    ($L_V=4, \Delta t=0.1, L_U=10, h_i=0,h_f=2$), $\mathcal T_{13\times 13}$ ($L_V=2, \Delta t=0.06, L_U=8, h_i=0, h_f=2$) and $\mathcal T_{\overline{30}\times \overline{30}}$ ($L_V=2, \Delta t=0.06, L_U=6, h_i=0, h_f=1.6$). c) compression of the Floquet Hamiltonian for two sizes of square periodic lattices, $\mathcal T_{\overline{8}\times \overline{8}}, \mathcal T_{\overline{10}\times \overline{10}}$. In both cases $L_V=2, \Delta t=\tau/7, L_U=8$. We observe that at odd periods the Trotterization has a very similar cost to the compressed ansatz, but that in general the Trotterization is very inaccurate, whereas the compressed ansatz outperforms it by many orders of magnitude.
    d) Improvement over Trotterization for different ramp speeds for the TFIM with identical choices of $h_i=0, h_f=2$ on the $\mathcal T_{11\times11}$ topology. Steeper ramps lead to shorter overall evolution time, and hence to larger improvement over Trotterization. e) Improvement over Trotterization for the data presented in panel b).}
    All choices of truncations are reported in \Cref{tab:PPtrunc} in SM.
    \label{fig:numsec}
\end{figure*}
\paragraph{Numerical results}
In this section we present the numerical results for compressing the dynamics of physical systems in two dimensions with the topology, $\mathcal T$. We study the dynamics generated by (i) the transverse field Ising model Hamiltonian with nearest-neighbor interactions, where the transverse field ramps linearly from the initial value $h_i$ to the final value $h_f$
\begin{equation}
    H_{\rm TFIM}(t') = \frac J2\sum_{\langle j,k\rangle_{\mathcal T}}Z_jZ_k - h(t')\sum_{j} X_j,
\end{equation}
with
\begin{equation}
    h(t') = (h_f-h_i) \frac{t'-t_i}{t_f-t_i} \equiv \Delta h \frac{t'-t_i}{t_f-t_i}
\end{equation}
and (ii) the Floquet Hamiltonian of the next-nearest-neighbor Ising model in a transverse field, where the strength of the transverse field is periodically modulated~\cite{eckstein2024large}
\begin{align}
    H_{\rm NNTFIM}(t')=&J\left(-\sum_{\langle j,k\rangle_{\mathcal T}}{Z}_j {Z}_k+\kappa \sum_{\langle\langle j,k\rangle\rangle_{\mathcal T}} {Z}_j {Z}_k\right) \nonumber\\ &-h \cos (\omega t') \sum_j {X}_j .
\end{align} 
Here, we use $\mathcal T_{n_x\times n_y}$ to indicate the connectivity of a rectangular lattice with ${n_x\times n_y}$ sites and open boundary conditions, where replacing $n_{x}, n_y$ by $\overline{n_{x}}, \overline{n_{y}}$ indicates periodic boundary conditions. We also consider  IBM's heavy-hex topology on $n$ qubits $\mathcal{T}_{n,\mathrm{h.-h.}}$. These different topologies are shown in \Cref{fig:numsec}a).
The notation $\langle j,k\rangle_{\mathcal T}$
and $\langle\langle j,k\rangle\rangle_{\mathcal T}$ indicates the locations of the nearest-neighbor and next-nearest-neighbor, respectively.

We target unitaries $U$ constructed from the time-ordered exponential $U(t)\approx \mathcal T\left\{\exp(i\int_{t_0}^t \mathrm dt'H(t'))\right\}$  using a fixed reference interval $\Delta t$ and approximating its dynamics with $L_U$ layers of a second order Trotterization; for a Hamiltonian $H$ with non-commuting terms $H=H_A+H_B$ this is given by $\mathcal T\left\{\exp(i\int_{t_0}^t \mathrm dt'H(t'))\right\}\approx \prod_{j=1}^{L_U} e^{-i \Delta t/(2L_U) H_A(t_j)}e^{-i \Delta t/L_U H_B(t_j)}e^{-i \Delta t/(2L_U) H_A(t_j)}$ ~\cite{trotter1959product}. In practice, we always used $\Delta t/L_U < 0.03$. The total evolution time $t$ is then reached by applying the $L_U$-layers circuit $t/\Delta t$ times.

\par The ansatz $V(\vec\theta)$ is also a second order Trotter circuit, except it consists of a smaller number of layers $L_V < L_U$, where the Hamiltonian has controllable parameters $\vec\theta$ associated with its interactions, e.g., $H=Z_1Z_2 + X_1\mapsto V(\vec\theta)=e^{i\theta_1 X_1}e^{i\theta_2Z_1Z_2}e^{i\theta_3 X_1}$ for $L_V=1$. 
This ansatz serves several purposes.
Firstly, as long as the total simulation time $t$ is not too long, choosing the initial parameters to correspond to the Trotter parameters, $\vec\theta_\mathrm{init} = \vec\theta_{\mathrm{trotter}}$, gives a physically motivated initial guess with non-vanishing gradients. This is crucial for large systems, where a random guess can potentially make the convergence of the optimization much more difficult.
Additionally, since our ansatz uses the same gates as the Trotterized circuit, the ability to simulate the dynamics of $U$ on a quantum processor also implies the ability to run the compressed ansatz $V$. 

We perform the compression of the dynamics of $H_{\mathrm{TFIM}}(t')$ and $H_{\rm NNTFIM}(t')$ for large $2$D systems. The results for $H_{\mathrm{TFIM}}$ and the topologies $\mathcal{T}_{127,\mathrm{h.-h.}}$, $\mathcal T_{13\times 13}$ and $\mathcal T_{\overline{30}\times \overline{30}}$ are in \Cref{fig:numsec}b). We observe that the optimized circuit obtained by minimizing $R_{\mathcal{Q}_{LS}}^{\mathrm{loc}}(\vec \theta^\star)$ can reduce the cost by a factor of $2\sim12$ compared to the Trotter circuit of the same depth and architecture with cost $R_{\mathcal{Q}_{LS}}^{\mathrm{loc}}(\vec \theta_{\mathrm{trotter}})$. Since PP of observables through deep circuits are usually very memory intensive, the use of translation invariance unlocks compressing very large systems, as exemplified by the compression on $\mathcal T_{\overline{30}\times\overline{30}}$. 
\begin{figure*}
    \centering
    \includegraphics[width=0.9\textwidth]{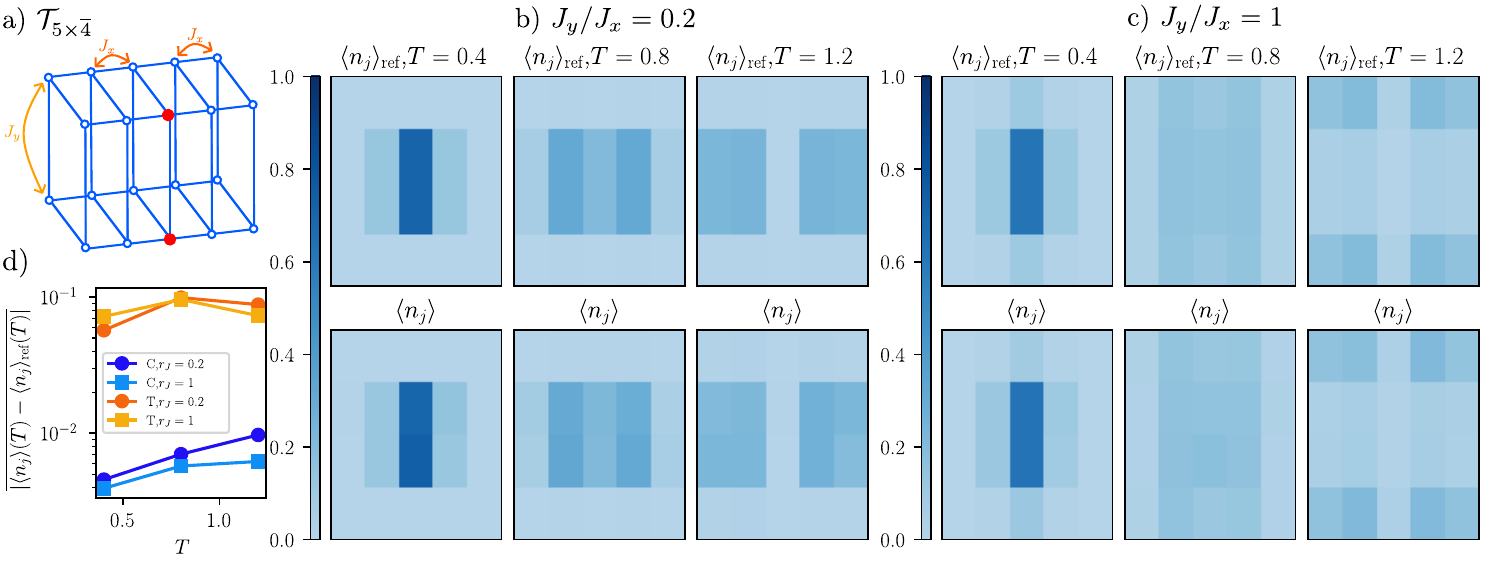}
    \caption{Dynamics of hard-core bosons with the compressed circuit on the H1 chip. a) Illustration of $\mathcal T_{5\times \overline 4}$. The two red dots correspond to the positions of the bosons at $T=0$, and we compress the dynamics for fixed $t=0.4$. b) H1 results for $r_J:=J_y/J_x=0.2$ c) H1 results for $r_J=J_y/J_x=1$. For both b-c): top row: expectation value of the occupation number at increasing times (from left to right) obtained from our exact statevector simulation. Bottom row: occupation number experimentally obtained by running the compressed circuit. d) Mean error in the occupation number for states prepared at different times using the compressed circuit and the Trotter circuit in real experiments. C stands for ``compressed'', T for ``Trotter''.}
    \label{fig:H1exp}
\end{figure*}
The results for the Floquet Hamiltonian with the topologies $\mathcal T_{\overline{8}\times \overline{8}}$, $\mathcal T_{\overline{10}\times \overline{10}}$ are in \Cref{fig:numsec}c). We observe that the Trotterization generally performs badly due to the time dependence of $H_{\rm NNTFIM}$ and its next-nearest neighbor interaction, as already remarked in \cite{eckstein2024large}. Specifically, we find $R_{\mathcal{Q}}^{\mathrm{loc}}(\vec\theta_\text{trotter})$ mostly stays above 0.01 for $t\gtrsim 2\tau$, where $\tau=2\pi/\omega$ is the period of the Floquet drive. On the other hand, the compressed ansatz performs well, and the cost remains between $10^{-6}$ and $10^{-4}$ for $t$ between $\tau$ and $3\tau$. Hence, our compressed ansatz can be up to $4$ orders of magnitude more accurate.

In \Cref{fig:numsec}d) we consider the ratio $R^{{\mathrm{{loc}}}}_{\mathcal{Q}_{LS}}(\vec\theta_{\mathrm{Trotter}})/R^{{\mathrm{{loc}}}}_{\mathcal{Q}_{LS}}(\vec\theta^\star)$, which quantifies the improvement over Trotter of the compressed ansatz, for different choices of $t_f-t_i$ at fixed $\Delta h$, hence for different ramp speeds. We consider the TFIM Hamiltonian on the $\mathcal T_{11\times11}$ topology. Starting from the time $(t-t_i)/(t_f-t_i)\approx 0.3$ we observe that steeper ramps lead to larger improvements. This is expected, as the steeper the ramp the shorter the time to reach $h_f$, and we expect that few-layers ansätze reproduce shorter time dynamics more accurately.

\paragraph{Experimental results}

\par For a quantum simulation on a device, one is interested in performing the dynamics for a long time $T=k\cdot t$ that cannot be simulated using classical methods. To achieve this, one can apply the $L_U$-layers unitary $U\approx\exp(-itH)$ multiple times, resulting in a circuit of $k\cdot L_U$ layers.
Since two-qubit operations in modern quantum hardware are still quite noisy, deep circuits tend to be inaccurate. To utilize the currently available quantum resources as effectively as possible, we are therefore motivated to find the shortest depth representation of the target dynamics.
\par In this section, we demonstrate the power of our compression algorithm by compressing the dynamics of a system of hard-core bosons placed on a $2$D lattice,
\begin{align}\label{eq:Hbos}
     H_{\mathrm{HCB}}=&-\frac{J_x}2\sum_{ \langle j,k\rangle_{\mathcal T|x}}\left[\hat{a}_{j}^{\dagger} \hat{a}_{k}+\hat{a}_{k}^{\dagger} \hat{a}_{j}\right] \nonumber\\
     &-\frac{J_y}2\sum_{ \langle j,k\rangle_{\mathcal T|y}}\left[\hat{a}_{j}^{\dagger} \hat{a}_{k}+\hat{a}_{k}^{\dagger} \hat{a}_{j}\right].
\end{align}
The notation $\langle j,k\rangle_{\mathcal T|x,y}$ denotes the pairs of sites that are connected by $\mathcal T$ in the $x$ or $y$ direction.
We consider the topology given by the $5\times 4$ cylinder $\mathcal T_{5\times \overline 4}$, see \Cref{fig:H1exp}a), and consider a target unitary with $L_U=12$ layers to be compressed into an ansatz with $L_V=2$ layers. The total compression time is chosen to be $t=0.4$. 
We perform the compression for two choices of ratios of the hopping strengths: $J_y/J_x=0.2$ and $J_y/J_x=1$.

By repeatedly applying the compressed circuit $k=1,2,3$ times to a product initial state with charge filled only at two sites at the center of the lattice (in the $x$ direction), see \Cref{fig:H1exp}a), we obtain the diffusion process at different evolution times on Quantinuum's H1 chip \cite{quantinuum2023}.
The results for the occupation number $\langle n_j\rangle$ are in \Cref{fig:H1exp}b-c). The distinction between the two diffusion processes is clearly captured by the compressed ansatz: for $J_y/J_x=0.2$ the diffusion is prevalently along the $x$ direction, whereas for $J_y/J_x=1$ we observe diffusion happening uniformly in the $x$ and $y$ directions. After post-selecting the output string for being in the correct charge sector (namely, having a total charge of 2), we observe good agreement between the results from the H1 chip and the reference values for $n_j$, as the average error rate $\overline{|\langle n_j\rangle(T) - \langle n_j\rangle_\mathrm{ref}(T)|}$ grows from $0.004$ at $T=0.4$ to $0.010$ at $T=1.2$. As a comparison, running the Trotter circuit on Quantinuum's H1 noisy emulator gives an average error of $0.056$ at $T=0.4$ and of $0.072$ at $T=1.2$ with the same data processing technique (see \Cref{fig:H1exp}d)). We also employ a noiseless statevector simulator to compute the fidelity of states prepared with either the compressed ansatz or standard Trotterization. States prepared with the compressed circuit have infidelities that are an order of magnitude better than states prepared with Trotter circuits (see \Cref{tab:bosons_extra} in Appendix). These results indicate that our compression algorithm is a useful tool for near-term devices by allowing for more accurate dynamical simulations than what would have been possible with standard Trotterization.  

\paragraph{Discussion}
By combining analytical derivations, numerical simulations, and experimental demonstrations, we have developed and demonstrated a strategy that enables compression of quantum circuits for large 2D system dynamics—a qualitative advance over existing  methods based on matrix product states, which remain confined to (quasi-)1D geometries or small 2D lattices.
We consistently observe order-of-magnitude accuracy improvements while preserving the original Trotterization architecture and gate sets, with particularly pronounced advantages for time-dependent Hamiltonians where standard Trotter methods degrade substantially~\cite{ikeda2023minimum}.

Several targeted extensions could enhance this approach. Generalizing to alternative gate sets, including direct compilation into hardware-native gates, would enable hardware-software co-design. Improved truncation schemes exploiting Hamiltonian symmetries or lightcone-insipired propagation could reduce the effective computational complexity by making calculations more efficient. This in turn would allow to compress deeper circuits.
Additionally, machine-learning techniques~\cite{carrasquilla2017machine,ch2017machine,torlai2018neural,amin2018quantum,fitzek2024rydberggpt} may offer substantial enhancements: learned initialization schemes, transfer learning across related Hamiltonians, or meta-learning strategies could dramatically reduce training time and improve convergence.
Ultimately, however, the fundamental question remains: what intrinsic physical features of a Hamiltonian—locality, symmetries, spectral properties, or dynamical structure—determine its compressibility and guide optimal compression strategies.

Our work advances practical quantum compilation by substantially reducing circuit depth requirements, bringing large-scale, long-time digital quantum simulations beyond classical capabilities within experimental reach.
\paragraph{Acknowledgements}
We would like to thank Dmitry Abanin, Alán Aspuru-Guzik,  Zoë Holmes, Hong-Ye Hu, Andrew Jreissaty, Yong-Baek Kim, Roger Luo, Roger G. Melko, Andrew Potter, Dvira Segal, and Yijian Zou for their insightful discussions.
YZ was supported by the Natural Science and Engineering Research Council (NSERC) of Canada and acknowledges support from the Center for Quantum Materials and Centre for Quantum Information and Quantum Control at the University of Toronto. MSR acknowledges funding from the 2024 Google PhD Fellowship and the Swiss National Science Foundation [grant number 200021-219329]. RW acknowledges support from the Flatiron Institute. The Flatiron Institute is a division of the Simons Foundation. Example simulation code used in this work can be found at~\url{https://github.com/MatteDAnna/cc2d_example}.

\bibliography{ref}
\clearpage

\appendix 
\onecolumngrid
\section{Methods}\label{sec:PP:appendix}
In this Method section, we detail the Pauli propagation (PP) method that is used throughout this work. PP is a computational framework for evolving observables $O$ expressed in the Pauli basis, under the action of a unitary $U$, $O\mapsto U^\dagger O U$, that has been used successfully to simulate noisy circuits \cite{schuster2024polynomialtimeclassicalalgorithmnoisy, fontana2025classical}, unstructured random circuits \cite{angrisani2024classicallyestimatingobservablesnoiseless} and dynamics in higher dimensions \cite{rudolph2023classical,Begu_i__2025}. PP addresses key limitations of Ref.~\cite{zhang2024scalable}, which demonstrated compression techniques for quasi-1D systems, allowing us to scale to fully 2D circuits. 

\subsection{Pauli Propagation framework}
PP is usually formulated in the Heisenberg picture; instead of considering the (forward) evolution of the initial state $\ket{\psi_0}$ under a unitary operator $U$ and then taking the expectation value of some observable $O$ for $\ket \psi = U \ket{\psi_0}$, $\bra\psi O\ket\psi$, we consider the backwards evolution of $O$ under $U^\dagger$, given by $O' = U^\dagger O U$, and then compute $\bra{\psi_0}O'\ket{\psi_0}$. However, \Cref{eq:locHaarpref} contains products of the form $\bbra{P_j} \bm{V(\vec\theta)U^\dagger} \kket{P_j}$, not overlaps against initial states $\bbra{\psi_0} \bm{V(\vec\theta)U^\dagger} \kket{P_j}$. We employ a ``meet in the middle'' approach and additionally consider the forward evolution of observables under $U$, $O'_f = U O U^\dagger$ to compute  $\bm{U^\dagger} \kket{P_j}$ and $\bbra{P_j} \bm{V(\vec\theta)}$ separately. In this work, we consider observables $O$ that can be written as
\begin{equation}\label{eq:OsumP}
    O = \sum_P a_P P,
\end{equation}
where $P\in \{I, X, Y, Z\}^{n}\equiv \mathbb P_n$ are the $4^n$ elements of the Pauli basis called Pauli strings, and $a_P\in\mathbb R$. To efficiently simulate the propagation, we require that the number of non-zero $a_p$ is $\mathcal O(\text{poly}(n))$ (or simply tractable). As a side note, Pauli propagation in the Schr\"odinger picture evolving pure states directly falls outside this constraint, as pure states $\rho$ always have exponentially many non-zero coefficients in the Pauli basis. 

We can formulate PP in the Pauli transfer matrix (PTM) formalism, where we take the Pauli basis elements as basis vectors, and we write observables as $4^n$ vectors. We denote the PTM representation of $O$ as $\kket{O}$, with entries
\begin{equation}
    \kket{O}_j = \tr(OP_j)/2^n = a_{P_j}.
\end{equation}
In this formalism, Pauli basis elements are sparse, $\kket{P_k}_j = \delta_{kj}/2^n$, and quantum states are often dense. For example, since $\ketbra 0 =((I+Z)/2)^{\otimes n}$, its PTM representation $\kket{0}$
has the $2^n$ combinations of $\{I,Z\}^{n}$ with non-zero coefficients. 
A unitary $U$ can be represented as the $4^n\times 4^n$ matrix $\mathbf U$ with entries 
\begin{equation}
    \mathbf U_{ij}= \bbra{P_i} \mathbf U \kket{P_j} = \tr\left(P_i U P_j U^\dagger\right)/2^n,
\end{equation}
although most unitaries of practical interest have structure that makes them representable in a sparser manner. Pauli rotations, for example, only require the computation of 4 terms (see Eq.~\eqref{eq:full_paulirot}) and Clifford gates are known to have only $\mathcal{O}(n^2)$ complexity to be applied~\cite{gottesman1998heisenberg}.

To this end, we can write quantum circuits as
\begin{equation}
    U(\vec\theta) = C_m e^{-\I\theta_m/2 P_{m}} C_{m-1} \cdots C_1 e^{-\I\theta_1/2 P_{1}} C_0,
\end{equation}
where each $C_j$ is a layer consisting only of Clifford gates and each $P_j$ is a Pauli basis vector. Note that the dynamics simulations performed in this work do not require any Clifford gates, but we are still giving a more general expression for completeness. Since Clifford gates are the stabilizers of the Pauli group, we have that for any $P_j\in\mathbb P_n$ and any $k=0,\cdots, m$
\begin{equation}\label{eq:CliffordProp}
    \mathbf{C_k} \kket{P_j} = c_k \kket{P_j'},
\end{equation}
where $\kket{P_j'}\in \mathbb P_n$, and  $c_k=\pm1$. 

The main focus in our work, however, are Pauli rotations that natively arise from the Trotterization of physical Hamiltonians. That is, we are interested in computing $\left(\bm{e^{-\I\theta/2 P_k}}\right)^T \kket{P_j}$ which has the form
\begin{equation}\label{eq:full_paulirot}
    e^{i \theta/2 P_k} P_j e^{-i \theta/2 P_k} = \left(\cos \left(\frac{\theta}{2}\right) I+i \sin \left(\frac{\theta}{2}\right) P_k\right)P_j \left(\cos \left(\frac{\theta}{2}\right) I-i \sin \left(\frac{\theta}{2}\right) P_k\right),
\end{equation}
and since Pauli strings either commute or anticommute we find, through trigonometric identities, the \textit{branching rule} under Pauli rotations
\begin{equation}\label{eq:PP_rotations}
    \bm{e^{-\I\theta/2 P_k}} \kket{P_j} = \begin{cases}\kket{P_j}, & \text { if }\left[P_k, P_j\right]=0 \\ \cos (\theta) \kket{P_j}\mp v_k\sin (\theta) \kket{P_k P_j}, & \text { if }\left\{P_k, P_j\right\}=0 .\end{cases}
\end{equation}
Here we used that $P_k P_j\in \mathbb P_n$ up to an imaginary factor $p_k=\pm \I$, with a slight abuse of notation, and $v_k=\I p_k=\pm1$. This means that if $P_j$ anticommutes with $P_k$, we now have two distinct Pauli strings, one with a prefactor of $\cos(\theta)$, the other with a prefactor of $\sin(\theta)$. We say that the Pauli string \textit{splits into two branches} (each branch with a different Pauli string). If a string splits, we assign its two branches the ``frequency label'' $\omega_k=\pm1$, whereas if it did not split $\omega_j=0$. We therefore define ``paths'' by the frequency vector $\vec\omega \in \{0,\pm1\}^m$. We introduce the trigonometric basis functions, indexed by all possible paths $\vec\omega$ as
\begin{equation}
    \Phi_{\vec\omega}(\vec{\theta}):=\prod_{i=1}^m c_iv_i\cdot \begin{cases}1 & \text { if } \omega_i=0 \\ \cos \left(\theta_i\right) & \text { if } \omega_i=1 \\ \sin \left(\theta_i\right) & \text { if } \omega_i=-1\end{cases},
\end{equation}
where $c_i=\pm1$ are the ``Clifford signs'' introduced in~\Cref{eq:CliffordProp} and $v_i=\pm1$ is the ``Pauli rotation sign'' defined as in~\Cref{eq:PP_rotations} for $\omega_i=-1$, whereas for $\omega_i=0,1$ we have $v_i=1$. The trigonometric basis functions allow us to expand $\mathbf U^\dagger \kket{P_j}$ into the modes $\vec\omega$
\begin{equation}
    \mathbf U^\dagger \kket{P_j} = \sum_{\vec\omega \in \Omega} \Phi_{\vec\omega}(\vec{\theta}) \kket{P_{\vec\omega}},
\end{equation}
where $\kket{P_{\vec\omega}}$ is the Pauli that results from starting in $P_j$ and following the path given by $\vec\omega$, and $\Omega\subseteq \{\{0,\pm1\}^m\}$ is the subset of all possible paths that are actually produced when propagating $P_j$ through the circuit.
Overlaps of the form $\bbra{P_j} \bm{V(\vec\theta)U^\dagger} \kket{P_j}$ can then by computed as via
\begin{equation}
    \bbra P_j \mathbf V \mathbf U^T \kket{P_j} = \sum_{\vec\omega_V\in \Omega}\sum_{\vec\omega\in \Omega} \Phi_{\vec\omega_V}\Phi_{\vec\omega}(\vec{\theta})\tr(P_{\vec\omega_V}P_{\vec\omega})/2^n,
\end{equation}
where $\vec\omega_V$ now equivalently enumerate the Pauli paths from the evolution under the target unitary $V$. This double sum is significantly more efficient than it may look because the only non-zero contributions in the double sum will be the products of coefficients to identical Pauli strings $P_{\vec\omega_V}$ and $P_{\vec\omega}$. As such, evaluation is linear in the shorter of the two sums by use of fast look-up data structures such as hash maps (e.g., dictionaries).

Importantly, we will use both the propagating Pauli strings and their coefficient, i.e., how many sine factors it has accumulated, to truncate to the dominant contributions.


\subsection{Truncation methods used in this work}\label{sec:PP:appendix:truncations}
Keeping track of all paths $\vec\omega$ quickly becomes computationally expensive, as their number grows exponentially in the number of layers $m$. It is therefore natural to employ some truncation strategy: at any point of the PP algorithm, we want to be able to decide if a certain path will contribute significantly or not. If not, we want to truncate that path. In this work, we employ three truncation schemes: coefficient truncation, weight truncation, and small-angle truncation.
\begin{figure*}
    \centering
    \includegraphics[width=.95\textwidth]{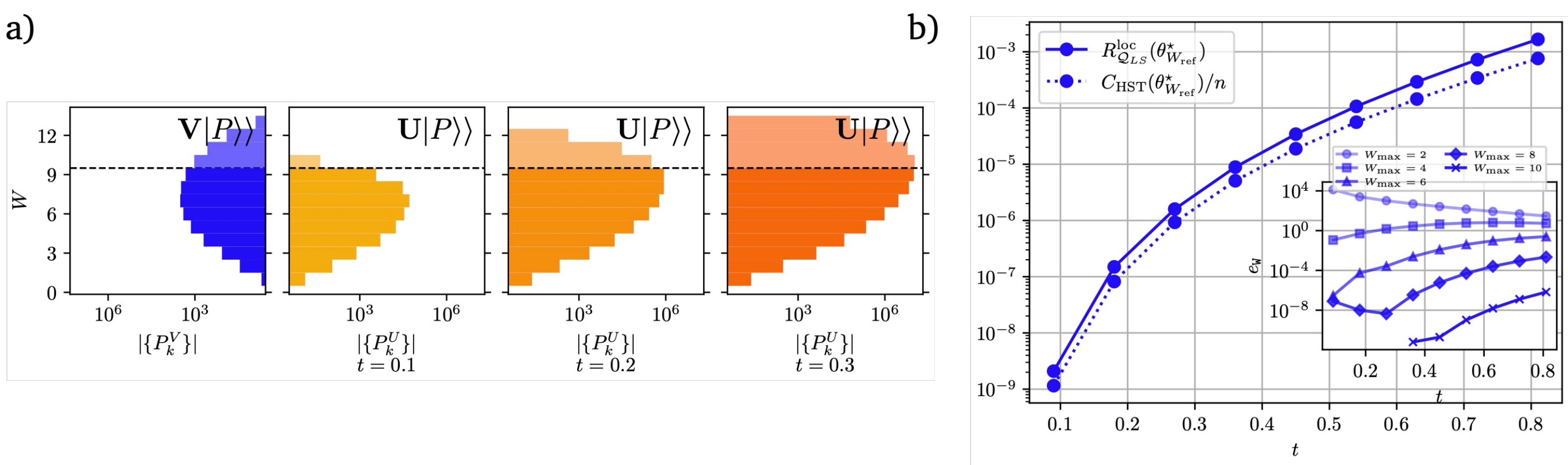}
    \caption{a) Growth of Pauli paths for $\bm {V}\kket{P}$ and $\bm U\kket P$ for different simulation times $t$. Here we illustrate weight truncation. Longer simulation times produce much bigger propagation trees. b) Comparison, at $\mathcal T_{\overline 4\times \overline 3}$, of $R^\mathrm{loc}_{\mathcal{Q}}$ and $C_\text{HST}$ for compressing the TFIM Hamiltonian. We observe that $R^\mathrm{loc}_{\mathcal{Q}}(\theta^\star_{W_\text{ref}})$ and $C_\text{HST}(\theta^\star_{W_\text{ref}})/n$ differ by a constant. Inset: relative error $e_{W} = \vert (R^\mathrm{loc}_{\mathcal{Q}}(\theta^\star_{W}) - R^\mathrm{loc}_{\mathcal{Q}}(\theta^\star_{W_\text{ref}}))/R^\mathrm{loc}_{\mathcal{Q}}(\theta^\star_{W_\text{ref}}) \vert$ for different weight truncations, with $W_\text{ref} = 12$. We cut off data for the $e_{W}$ inset at $10^{-15}$. }
    \label{fig:truncations}
\end{figure*}

\par We define the weight $W$ of a Pauli string $P$ as the number of components it is acting on nontrivially. Note that $\langle \langle 0|P_{\vec\omega}\rangle\rangle\neq 0$ if $P_{\vec\omega}$ only contains $I, Z$ because
\begin{align*}
    \langle\langle{0} \kket{ P_{\vec\omega}} &= \tr(\ketbra{0} P_{\vec\omega})/2^n \\
    &= \tr\left(\left(\frac{1}{2}(I+Z)\right)^{\otimes n} P_{\vec\omega}\right)/2^n
\end{align*}
is only non-zero for paths that do not contain $X$ or $Y$. 
Strings of higher weight are more likely to contain at least one $X,Y$, which means they are more likely not to contribute. The \textit{weight-based truncation} strategy corresponds to setting a maximum allowed weight $W_\mathrm{max}$, and truncating all paths that at any moment have weight higher than said maximum weight.

\par The \textit{coefficient truncation} method is about the numerical prefactor to each string: if at any point of the circuit this numerical coefficient $a_P$ has absolute value smaller than a fixed cutoff value $\epsilon$, $\vert a_P\vert < \epsilon$, the path is truncated.

\par Another useful truncation method is the \textit{small angle truncation}: if the gate angles satisfy $|\theta_j| \ll 1$, $\forall j$ (which is typical for dynamics), then $|\sin(\theta_j)| \approx |\theta_j| \ll 1$. For a path $\vec\omega$, let 
\begin{equation}
        r_\pm(\vec\omega):= \sum_{i=1}^m \delta_{\pm1, \omega_i},
\end{equation}
be the number of modes with $+1$ and $-1$, respectively. In the small angle approximation $\theta_j \approx \theta \ll 1$, we have
\begin{equation}
    \abs{\Phi_{\vec\omega}(\vec{\theta})} \approx \theta^{r_-(\vec\omega)},
\end{equation}
which implies that the contribution of a path rapidly decreases with the number of $\sin$ modes it accumulates. Hence, one can expand the propagation in orders of the number of sines kept and truncate all paths with $r_-(\vec\omega)$ above a threshold $r_-^{\mathrm{max}}$. In \cite{lerch2024efficientquantumenhancedclassicalsimulation}, it was found that this truncation is provably efficient in approximating expectation values to constant additive error when Pauli rotation angles are of magnitude $\mathcal{O}(1/m)$, where $m$ is the number of Pauli rotations. This scaling is interestingly valid for average-case errors in circuits with correlated angles (like Trotter circuits) and worst-case circuits. Average-case simulation of circuits with uncorrelated angles is efficient with larger angle ranges.

The small angle truncation is a ``sub-type'' of coefficient truncation. A big advantage that this truncation has over pure coefficient truncation is about the optimization process. In this work we need to extract the gradients of our cost function, and the the best-performing AD library we used was \texttt{ReverseDiff.jl} and its prerecorded gradient tapes. Because of this, we could not truncate based on coefficients (as different choices of $\vec\theta$ have different branches truncated out because of coefficient truncation), but it was possible to truncate on $r_-$.

\par The small angle truncation is particularly useful for evaluating PTM elements of the form $\bbra{P_j} \bm{V(\vec\theta)U^\dagger} \kket{P_j}$, where, according to \Cref{eq:PP_rotations}, a Pauli string is modified only when a $\sin$ mode is encountered. Therefore, $r_-$ can be interpreted as a measure of how much a string has been modified. From a dynamical perspective, $r_-$ can also be interpreted as a ``lightcone truncation'': modified strings with large $r_-$ lie at the edge of the lightcone generated by $\bm{V(\vec\theta)}\kket{P_j}$ and often fall outside the lightcone of $\bm{U^\dagger} \kket{P_j}$, which is typically much deeper (this less technical point of view was chosen as the motivation for the small angle truncation in the main text). The expectation value is then computed as $(\bbra{P_j}\mathbf V) \cdot (\bm{U^\dagger}\kket{P_j})$, with Pauli strings being propagated both forward through $V(\vec\theta)$ and backward through $U$ instead of propagating through $V(\vec\theta)U^\dagger$ in one go. This ``meet in the middle'' approach heavily improves the computation, as only a small fraction of the paths generated by $\bm{VU^\dagger}\kket{P_j}$ have an overlap with $\bbra{P_j}$, and such paths typically have a small value of $r_-$, motivating truncating higher $r_-$. While this is a heuristic argument (since one must also consider the growing number of such paths), in practice, truncating high-$r_-$ paths has proven very effective in discarding non-contributing terms. 

Furthermore, if we consider the two-qubit Pauli rotations, only the sine branches can increase the Pauli weight (and generally tend to do so). Thus, high $r_-$ paths not only have smaller coefficients, but also tend to have higher weight. This accelerates the accumulation of more sine coefficients, but importantly also directly affects the inner product of the two half-evolved Pauli sums, where the expectation values are dominated by Pauli strings with lower weight.

\par Finally, one technical caveat is that strings with high $r_-$ could, in principle, propagate back toward the lightcone center and still contribute. Fortunately, \texttt{PauliPropagation.jl} uses a ``merging'' strategy that combines Pauli strings with different $r_-$ and propagates them as one with the smaller of the two sine counts. This implies that some paths that flow back earlier in the propagation will be kept for free despite them formally having accumulated more sine coefficients.

\begin{figure}
    \centering
    \includegraphics[width=0.95\linewidth]{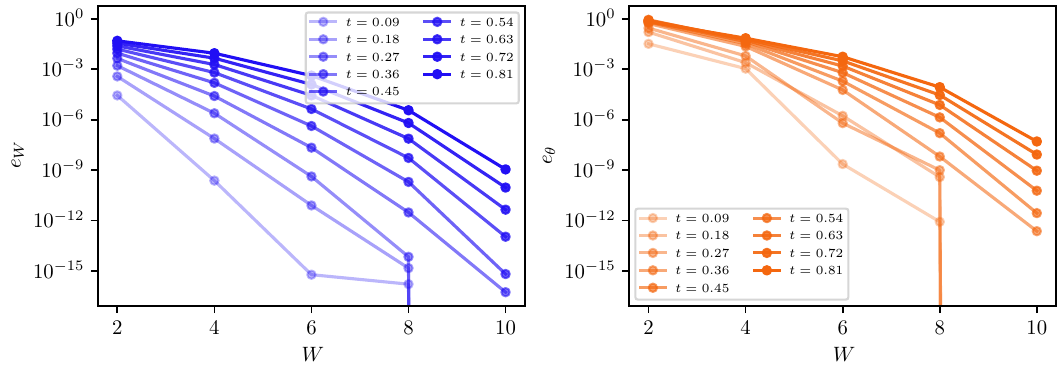}
    \caption{As in \Cref{fig:truncations} we consider the TFIM Hamiltonian with the topology $\mathcal T_{\overline4\times\overline3}$. Left: convergence of $e_W$ as a function of $W$ at different times. Right: convergence in $W$ of $e_\theta$ at different times.}
    \label{fig:errors}
\end{figure}
\subsection{Numerical justification}
To justify our compression method, we perform the compression of the dynamics generated by $H_{\mathrm{TFIM}}$ for a $4\times3$ system as accurately as possible, without imposing any weight truncation, $W_\text{max}\equiv W_\text{ref}=12$. We compare the results of $R^\mathrm{loc}_{\mathcal{Q}_{LS}}(\vec\theta^\star_{W_\text{ref}})$ with $C_{\mathrm{HST}}(\vec\theta^\star_{W_\text{ref}})$, since the Hilbert-Schmidt cost can be computed exactly at this system size.
As shown in \Cref{fig:truncations} b), $R^\mathrm{loc}_{\mathcal{Q}_{LS}}(\vec\theta^\star_{W_\text{ref}})$ and $C_\text{HST}(\vec\theta^\star_{W_\text{ref}})/n$ are very close to each other for all $t$. This indicates that our cost function $R^\mathrm{loc}_{\mathcal{Q}_{LS}}(\vec\theta)$ is a faithful proxy for the true cost $C_{\mathrm{HST}}(\vec\theta)$.
In the inset, we show the convergence of $R^\mathrm{loc}_{\mathcal{Q}_{LS}}(\vec\theta^\star_{W_\text{max}})$ for different weight truncations $W_\text{max}$, and we find that for $W_\text{max}=8$, we can reach a satisfactory convergence at all times.
We observe this across all the simulations we performed, with more demanding simulations reaching good convergence for $W_\text{max}=10, 12$.

In Table \ref{tab:PPtrunc} we report the numerical choices for all PP truncations we enforced to obtain the results presented in the main text.
\begin{table}
    \centering
    \begin{tabular}{ccccccccc}
    \toprule
    Figure & Model & Physical params & $\mathcal T$ & $W$ & $\epsilon_U$ & $\epsilon_V$ & $\max_{\sin,U}$ & $\max_{\sin,V}$\\
    \midrule 
    4 & TFIM & $J=1,h=1.1$ & $\mathcal T_{\overline 4\times\overline 3}$  & $12$ & $1\cdot 10^{-11}$ & $1\cdot 10^{-11}$ &  $16$ & $16$\\
    \midrule
    2b, 2e & TFIM & $J=1, h_i=0, h_f=2$ & $\mathcal T_{13\times 13}$  & 10 & $1\cdot 10^{-10}$ & 0 & 10 & 10 \\
    2b, 2e & TFIM & $J=1,h_i=0, h_f=1.6$ & $\mathcal T_{\overline{30}\times \overline{30}}$ & 10 & $1\cdot 10^{-10}$ & $1\cdot 10^{-11}$ & 12 & 12 \\
    2b, 2e & TFIM & $J=1, h_i=0, h_f = 2$& $\mathcal T_{127, h-h}$ & 11 & $1\cdot 10^{-11}$ & $1\cdot 10^{-12}$ & 12 & 11 \\
    \midrule 
    2c & NNTFIM & $J=1,h=1.5,\kappa=1.5, \omega=30$ & $\mathcal T_{\overline{8}\times \overline{8}}$  & 9 & $1\cdot 10^{-10} $& $1\cdot 10^{-11}$ & 12 & 12 \\
    2c & NNTFIM & $J=1,h=2.5,\kappa=1.5, \omega=30$& $\mathcal T_{\overline{10}\times \overline{10}}$  & 8 & $1\cdot 10^{-10}$ & $1\cdot 10^{-11}$ & 12 & 12 \\
    \midrule  
    2d, $t_f-t_i=0.16$ & TFIM & $J=1, h_i=0, h_f=2$ & $\mathcal T_{11\times 
    11}$ & 10 & $1\cdot 10^{-10}$ & 0 & 10 & 10 \\
    2d, $t_f-t_i=0.32$ & TFIM & $J=1, h_i=0, h_f=2$ & $\mathcal T_{11\times 
    11}$ & 10 & $1\cdot 10^{-10}$ & 0 & 10 & 10 \\
    2d, $t_f-t_i=0.64$ & TFIM & $J=1, h_i=0, h_f=2$ & $\mathcal T_{11\times 
    11}$ & 10 & $1\cdot 10^{-10}$ & 0 & 10 & 10 \\
    2d, $t_f-t_i=0.96$ & TFIM & $J=1, h_i=0, h_f=2$ & $\mathcal T_{11\times 
    11}$ & 10 & $1\cdot 10^{-10}$ & 0 & 10 & 10 \\
    \bottomrule 
    \end{tabular}
    \caption{Physical parameters and truncations for the plots presented in the main text. }
    \label{tab:PPtrunc}
\end{table}
\subsection{Monitoring convergence}
We monitor two errors to certify convergence: the relative error in $R^\mathrm{loc}_{\mathcal{Q}}$: $e_W = \vert (R^\mathrm{loc}_{\mathcal{Q}}(\theta^\star_W) - R^\mathrm{loc}_{\mathcal{Q}}(\theta^\star_{W_\text{ref}}))/R^\mathrm{loc}_{\mathcal{Q}}(\theta^\star_{W_\text{ref}}) \vert$ and the difference between the optimal parameters obtained at different weight truncations:
$e_\theta(W) = || \vec\theta^\star_{W}-\vec\theta^\star_{W_\text{ref}}||$. Both $e_W,e_\theta$ for the same data presented in \Cref{fig:truncations} in the main text are in \Cref{fig:errors}. As expected, both $e_W,e_\theta$ converge more slowly for larger times, but nonetheless reach good convergence.

Further details on PP are provided in the SM and Ref.~\cite{rudolph2025pauli}.

\section{Analytical deriviations}
In this section, we give the details needed to derive \Cref{eq:locHaarpref}. 
We derive the exact average for sampling states from locally scrambling ensembles by using the single-qubit Haar states ensemble as an explicit example. To introduce the notation, we start with the case of $n=1$, which is then generalized to the case for any $n$.

\subsection{Explicit results for the single qubit Haar measure}
\begin{figure*}
    \centering
    \includegraphics[width=0.7\linewidth]{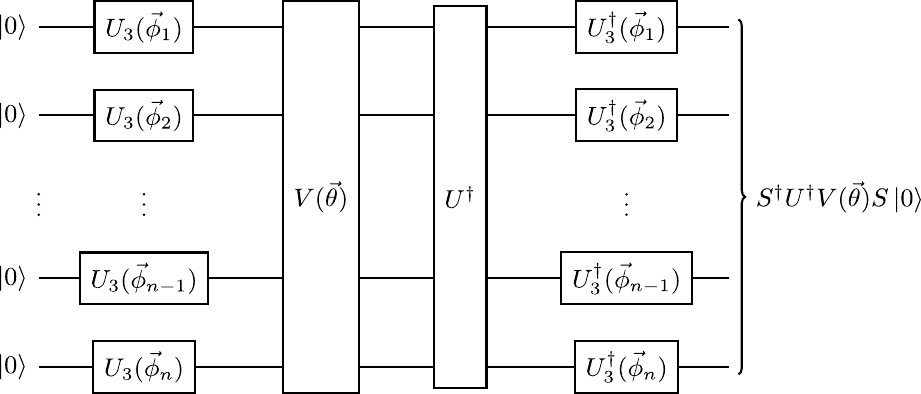}
    \caption{Circuit needed to compute the expectation values $\mathbb E \left[\langle 0| L^\dagger Z_j L|0 \rangle  \right]_{S \sim \mathcal Q}$}
    \label{fig:fullcirc}
\end{figure*}
In the context of VQC, we are interested in the expectation values in~\Cref{eq:Rloc},
\begin{align}
    \mathbb E\left[\bbra{0} \bm {L_S(\vec\theta)} \kket{Z_j}\right]_{S \sim \mathcal{Q}_{LS}} = \mathbb E \left[\langle 0| L_S(\vec \theta) Z_j L_S(\vec \theta)^\dagger|0 \rangle  \right]_{S \sim\mathcal{Q}_{LS}} ,\label{eq:exp_val}
\end{align}
where $\mathcal{Q}_{LS}$ is a locally scrambling ensemble and $L_S(\vec \theta):=S^\dagger U^\dagger  V(\vec \theta) S$. Remember that we have 
\begin{align}
    \bbra{P_i} \mathbf{W}\kket{P_j} = \tr\left[P_i W P_j W^\dagger\right], \label{eq:trace_U}
\end{align}
hence we get vanishing contributions if the Paulis generated by $\bbra{0}\mathbf{S^\dag}$ are orthogonal to those generated by $\mathbf{U^\dag}\mathbf{V}\mathbf{S}\kket{Z_j}$. 

An example of a locally scrambling ensemble is the ensemble $\mathrm{Haar}_{1}^{\otimes n}$, which corresponds to the tensor product of single qubit Haar random unitaries. Since our results are independent of the choice of locally scrambling ensemble, we can set $\mathcal{Q}_{LS}=\mathrm{Haar}_{1}^{\otimes n}$ and calculate the expectation value of \Cref{eq:exp_val}.

To generate unitaries according from $\mathrm{Haar}_{1}^{\otimes n}$, we can apply a random single qubit unitary acting to each qubit $j$,
\begin{align}
    U_3(\vec\phi):=U_3(\varphi,\gamma,\omega) = R_Z(\varphi)R_Y(\gamma)R_Z(\omega),\label{eq:U3}
\end{align}
where the angles are sampled from the appropriate measure (see below). In the PTM picture, this gives
\begin{align}
    \mathbf{U}_3(\vec\phi) \kket{Z_j} &= \mathbf{R_Z}(\varphi)\mathbf{R_Y}(\gamma)\mathbf{R_Z}(\omega)\kket{Z_j}\nonumber\\
    &=\mathbf{R_Z}(\varphi)
    \mathbf{R_Y}(\gamma)\kket{Z_j}\nonumber\\
    &= \mathbf{R_Z}(\varphi)\left(\cos(\gamma)\kket{Z_j} + \sin(\gamma)\kket{X_j}\right)\nonumber\\
    &= \cos(\gamma)\kket{Z_j} + \cos(\varphi)\sin(\gamma)\kket{X_j}+\sin(\varphi)\sin(\gamma)\kket{Y_j} \\
    & =:  \Phi_Z(\vec\phi) \kket{Z_j} + \Phi_X(\vec\phi) \kket{X_j}  + \Phi_Y(\vec\phi) \kket{Y_j}\label{eq:U3Z}
\end{align}

\subsection{Single qubit}
For a single qubit, we have take $\kket{Z_j}\equiv \kket{Z}$ and
\begin{align*}
    \bbra{0} = \frac{1}{2}(I + Z).
\end{align*}
Note that due to \Cref{eq:trace_U} 
\begin{align}
    \bbra{I}\mathbf{S}^\dag \mathbf{U^\dag}\mathbf{V}(\vec\theta)\mathbf{S}\kket{Z} = 0,\label{eq:zero_id}
\end{align}
hence we have to compute 
\begin{equation}
    \bbra{0}\mathbf{S}^\dag \mathbf{U^\dag}\mathbf{V}(\vec\theta)\mathbf{S}\kket{Z} = \bbra{Z}\mathbf{S}^\dag \mathbf{U^\dag}\mathbf{V}(\vec\theta)\mathbf{S}\kket{Z}/2.
\end{equation}
We have to apply the single qubit rotation of \Cref{eq:U3} to $\kket{Z}$ and $\bbra Z$. For $\kket{Z}$ we have \Cref{eq:U3Z}.
For $\bbra{Z}\mathbf{U}_3(\vec\phi)^\dag$ we find the analogous expression
\begin{align}
    \bbra{Z}\mathbf{S}^\dag = \bbra{Z}\mathbf{U}_3(\vec\phi)^\dag= \cos(\gamma)\bbra{Z}  + \cos(\varphi)\sin(\gamma)\bbra{X}+\sin(\varphi)\sin(\gamma)\bbra{Y},\label{eq:U3dagZ}
\end{align}
by noticing that $U_3\to U_3^\dag$ implies $\vec\phi\to-\vec\phi$, but also $\kket Z\to\bbra Z$ implies $\vec\phi\to-\vec\phi$, so the two effect cancel out and there are no additional signs.
Under the uniform measure 
\begin{equation}
    \mathbb{E}[A]_{\mathrm{Haar}_1} := \int \mathrm d\mu(\vec\phi) :=  \frac{1}{2(2\pi)^2}\int_{0}^{2\pi} \mathrm d\varphi\int_{0}^{2\pi}\mathrm d\omega \int_0^\pi \sin(\gamma)\mathrm d\gamma A,
\end{equation}
 we then find, by plugging in the results of Equations \ref{eq:U3Z} and \ref{eq:U3dagZ},
\begin{align*}
    \mathbb E  \left[\bbra{Z} \mathbf{S}^\dag \mathbf{U^\dag}\mathbf{V}(\vec\theta)\mathbf{S}\kket{Z}\right]_{S \sim \mathrm{Haar}_1} 
    &= a_{XX}(\varphi, \gamma)\bbra{X} \mathbf{U^\dag}\mathbf{V}(\vec\theta)\kket{X}  
    + a_{XY}(\varphi, \gamma)\bbra{X} \mathbf{U^\dag}\mathbf{V}(\vec\theta)\kket{Y} 
    + a_{XZ}(\varphi, \gamma)\bbra{X} \mathbf{U^\dag}\mathbf{V}(\vec\theta)\kket{Z}\\
    & +a_{YX}(\varphi, \gamma) \bbra{Y}\mathbf{U^\dag}\mathbf{V}(\vec\theta)\kket{X}
    + a_{YY}(\varphi, \gamma)\bbra{Y} \mathbf{U^\dag}\mathbf{V}(\vec\theta)\kket{Y} 
    +a_{YZ}(\varphi, \gamma)\bbra{Y} \mathbf{U^\dag}\mathbf{V}(\vec\theta)\kket{Z}\\
    & + a_{ZX}(\varphi, \gamma)\bbra{Z}\mathbf{U^\dag}\mathbf{V}(\vec\theta)\kket{X} 
    + a_{YX}(\varphi, \gamma)\bbra{Z} \mathbf{U^\dag}\mathbf{V}(\vec\theta)\kket{Y} 
    + a_{ZZ}(\varphi, \gamma)\bbra{Z}\mathbf{U^\dag}\mathbf{V}(\vec\theta)\kket{Z},
\end{align*}
where we have
\begin{align*}
    a_{XX}(\varphi, \gamma) & :=\mathbb{E}[\Phi_X^2]_{\mathrm{Haar}_1}  = \frac{1}{2(2\pi)^2}\int_{0}^{2\pi} \mathrm d\varphi(\cos(\varphi))^2\int_{0}^{2\pi}\mathrm d\omega \int_0^\pi \mathrm d\gamma (\sin(\gamma))^3 = \frac{1}{3}\\
    a_{YY}(\varphi, \gamma) &:=\mathbb{E}[\Phi_Y^2]_{\mathrm{Haar}_1}= \frac{1}{2(2\pi)^2}\int_{0}^{2\pi} \mathrm d\varphi (\sin(\varphi))^2 \int_{0}^{2\pi}\mathrm d\omega \int_0^\pi \mathrm d\gamma(\sin(\gamma))^3 =  \frac{1}{3}\\
    a_{ZZ}(\varphi, \gamma) &:=\mathbb{E}[\Phi_Z^2]_{\mathrm{Haar}_1}= \frac{1}{2(2\pi)^2}\int_{0}^{2\pi} \mathrm d\varphi\int_{0}^{2\pi}\mathrm d\omega \int_0^\pi \mathrm d\gamma \sin(\gamma)(\cos(\gamma))^2 = \frac{1}{3}\\
    a_{XY}(\varphi, \gamma) \equiv a_{YX}(\varphi, \gamma) &:=\mathbb{E}[\Phi_X\Phi_Y]_{\mathrm{Haar}_1} = \frac{1}{2(2\pi)^2}\int_{0}^{2\pi} \mathrm d\varphi \cos(\varphi)\sin(\varphi)
    \int_{0}^{2\pi}\mathrm d\omega \int_0^\pi (\sin(\gamma))^3\mathrm d\gamma
    = 0\\
    a_{YZ}(\varphi, \gamma) \equiv a_{ZY}(\varphi, \gamma) &:=\mathbb{E}[\Phi_Y\Phi_Z]_{\mathrm{Haar}_1}= \frac{1}{2(2\pi)^2}\int_{0}^{2\pi} \mathrm d\varphi \sin(\varphi) \int_{0}^{2\pi}\mathrm d\omega \int_0^\pi \mathrm d\gamma
    (\sin(\gamma))^2\cos(\gamma) = 0\\
    a_{XZ}(\varphi, \gamma) \equiv a_{ZX}(\varphi, \gamma) & :=\mathbb{E}[\Phi_X\Phi_Z]_{\mathrm{Haar}_1}= \frac{1}{2(2\pi)^2}\int_{0}^{2\pi} \mathrm d\varphi \cos(\varphi)
    \int_{0}^{2\pi}\mathrm d\omega \int_0^\pi \mathrm d\gamma
    (\sin(\gamma))^2\cos(\gamma) = 0,
\end{align*}
Where we used the trigonometric identities
\begin{align}
    \int_{0}^{2\pi} \mathrm d\varphi \cos(\varphi)\sin(\varphi) &= \int_{0}^{2\pi} \mathrm d\varphi \sin(\varphi)=\int_{0}^{2\pi} \mathrm d\varphi \cos(\varphi)=0 \label{eq:trig1}
\end{align}
to see that all cross terms are zero, while for $a_{XX}(\varphi, \gamma)$, $a_{YY}(\varphi, \gamma)$ and $a_{ZZ}(\varphi, \gamma)$ we used that
\begin{align}
    \int_{0}^{2\pi} \mathrm d\varphi(\cos(\varphi))^2 &= \int_{0}^{2\pi} \mathrm d\varphi(\sin(\varphi))^2 = \pi \nonumber \\
    \int_{0}^{2\pi} \mathrm d\varphi \cos(\varphi)\sin(\varphi) &= \int_{0}^{2\pi} \mathrm d\varphi \sin(\varphi)=\int_{0}^{2\pi} \mathrm d\varphi \cos(\varphi)=0 \nonumber\\
    \int_0^\pi \mathrm d\gamma (\sin(\gamma))^3 & =  \frac{4}{3} \nonumber \\
    \int_0^\pi \mathrm d\gamma (\cos(\gamma))^2 \sin(\gamma) & = \frac{2}{3}.\label{eq:trig2}
\end{align}
All the above identities can be summarized as
\begin{equation}\label{eq:ortho}
    \mathbb{E}[\Phi_P\Phi_Q]_{\mathrm{Haar}_1} = \frac{\delta_{P,Q}}3.
\end{equation}
for $P,Q\in\{X,Y,Z\}$. Combined with Equation \ref{eq:zero_id} we finally obtain
\begin{align*}
    \mathbb E  \left[\bbra{0} \mathbf{U^\dag}\mathbf{V}(\vec\theta)\kket{Z}\right]_{S \sim \mathrm{Haar}_1} 
    &= \frac{1}{3}\left(\bbra{X} \mathbf{U^\dag}\mathbf{V}(\vec\theta)\kket{X} +\bbra{Y} \mathbf{U^\dag}\mathbf{V}(\vec\theta)\kket{Y}+\bbra{Z} \mathbf{U^\dag}\mathbf{V}(\vec\theta)\kket{Z}\right).
\end{align*}

\subsection{Multiple qubits}
For multiple qubits, we see that the zero states becomes a linear combination of an exponential number of states
\begin{align*}
    \ketbra{0} = \frac{1}{2^n}(I + Z)^{\otimes n} = \frac{1}{2^n}\sum_b Z_b,
\end{align*}
where we defined
\begin{align}
    Z_b := Z^{b_0}\otimes Z^{b_1}\otimes \ldots \otimes  Z^{b_n},\label{eq:Zb}
\end{align}
with $b\in \{1,2\}^n$ and $Z^2=I$ which gives $2^n$ possible Pauli strings $Z_b$. Now we will show that the only non-zero contribution to Equation \ref{eq:exp_val} arises from the term $\bbra{Z_j}$, with all other $\bbra{Z_b}$ giving a vanishing contribution. 
To start, we see that backpropagating the $\bbra{Z_b}$ operators through $\mathbf{S}^\dag$ will in general result in a linear combination of all Pauli strings $P_b\in\mathbb{P}_n$ since $\mathbf{S}^\dag$ can scramble $Z^{\otimes n}$ to a linear combination of all $P_b\in\mathbb{P}_n$. 
Therefore the expectation value $\bbra{0} \mathbf L^T\kket{Z_j}$ can be expanded as
\begin{equation}
    \bbra{0} \mathbf L^T\kket{Z_j} = \sum_{P_b\in\mathbb P_n} \sum_{Q=X,Y,Z} \left( \bbra{P_b} \mathbf{UV^T}\kket{Q_j} \Phi_Q(\vec\phi_j)\prod_{i=1}^n \Phi_{P_{b_i}}(\vec\phi_i)\right),
\end{equation}
by extending the notation to $\Phi_I(\vec\phi):=1$. Therefore
\begin{equation}
    \begin{split}\mathbb E \left[\bbra{0} \mathbf L^T\kket{Z_j}  \right]_{S \sim \mathcal U_{\text{Haar}_1^{\otimes n}}} &= \int\cdots\int 
    \mathrm d\mu(\vec\phi_1)\cdots \mathrm d\mu(\vec\phi_n)\sum_{P_b\in\mathbb P_n} \sum_{Q=X,Y,Z} \left( \bbra{P_b} \mathbf{UV^T}\kket{Q_j} \Phi_Q(\vec\phi_j)\prod_{i=1}^n \Phi_{P_{b_i}}(\vec\phi_i)\right)\\
    &= \sum_{P_b\in\mathbb P_n} \sum_{Q=X,Y,Z}\bbra{P_b} \mathbf{UV^T}\kket{Q_j}\int\cdots\int 
    \mathrm d\mu(\vec\phi_1)\cdots \mathrm d\mu(\vec\phi_n) \Phi_Q(\vec\phi_j)\prod_{i=1}^n \Phi_{P_{b_i}}(\vec\phi_i).\\
    \end{split}
\end{equation}
We start by extending \Cref{eq:ortho} for the multi-qubit case, noticing that single-qubit expectations are vanishing if $Q=I$:
\begin{equation}
    \mathbb{E}[\Phi_P]_{\mathrm{Haar}_1} = \delta_{P,I},
\end{equation}
and therefore for integral over all sites

\begin{equation}
\begin{split}
    \int\cdots\int 
    \mathrm d\mu(\vec\phi_1)\cdots \mathrm d\mu(\vec\phi_n) \Phi_Q(\vec\phi_j)\prod_{i=1}^n \Phi_{P_{b_i}}(\vec\phi_i) &= \left(\int \mathrm d\mu(\vec\phi_j)\Phi_{P_{b_j}}(\vec\phi_j)\Phi_Q(\vec\phi_j)\right)\cdot \prod_{i\neq j}\left(\int \mathrm d\mu(\vec\phi_i)\Phi_{P_{b_i}}(\vec\phi_i)\right)\\
    & =\mathbb{E}[\Phi_{Q}\Phi_{P_{b_j}}]_{\mathrm{Haar}_1}\cdot \prod_{i\neq j}\mathbb{E}[\Phi_{P_{b_i}}]_{\mathrm{Haar}_1}\\
     & = \frac{\delta_{Q,P_{b_j}}}3\cdot \prod_{i\neq j}\delta_{I, P_{b_i}}
    \end{split}
\end{equation}

The expectation value~\Cref{eq:exp_val} hence becomes
\begin{equation}
    \mathbb E \left[\langle 0| L^\dagger Z_j L|0 \rangle  \right]_{S \sim \mathcal U_{\text{Haar}_1^{\otimes n}}}  = \frac 13 \langle\langle Z_j \vert \mathbf{VU^\dagger}\vert Z_j\rangle \rangle  +\frac 13 \langle\langle X_j \vert \mathbf{VU^\dagger}\vert X_j\rangle \rangle +\frac 13 \langle\langle Y_j \vert \mathbf{VU^\dagger}\vert Y_j\rangle \rangle
\end{equation}
\subsection{Advantages over sampling}
The standard approach to taking expectation values $E\left[\bullet\right]_{S \sim \mathcal{Q}_{LS}}$ with respect to ensembles of states $\mathcal{Q}_{LS}$ is to sample $M$ of them, and then approximate the expectation value as the average over those $M$. In this work we showed how to compute the expectation value exactly:
\begin{equation}  \lim_{M\to\infty} \frac 1M\sum_{S\sim\mathcal{Q}_{LS}}\bbra{0} \bm {L_S(\vec\theta)} \kket{Z_j} = \frac13\sum_{P=X,Y,Z}\langle\langle P_j| \cdots | P_j\rangle\rangle,\end{equation}
which is the sum of only of PTM diagonal entries, that is matrix elements of the form 
\[ \langle\langle P_j \vert \bm{V(\vec\theta)U^\dagger}\vert P_j\rangle \rangle.\]
The above PTM diagonal matrix elements are easier to evaluate, compared to the expectation value of the form $\bbra{0} \bm {L_S(\vec\theta)} \kket{Z_j}$, thanks to the ``meet in the middle'' approach and small angle truncation presented in the Truncations Section. This is because, since a Pauli string $\kket{P_j}$ gets modified only when it picks up a $\sin$ mode, see~\Cref{eq:PP_rotations}, the more of them it picks up the more the string gets modified. Hence, it becomes unlikely that such highly modified string will get back to the initial string $P_j$ by the end of the backpropagation and thus have a non-zero overlap with $\bbra{P_j}$. Although this is just a hand-waving argument because one has to take into account both the shrinking probability of contributing and the growing number of paths, in practice, truncating paths above a certain number of sin modes has proven very effective in discarding non-contributing paths.

\par This is even on top of the standard motivation for sin truncations, that the more sin modes a path picks up the smaller its contribution becomes. Consider for example a path $\vec\omega$ that reaches maximal weight $W$. Given that we only consider Hamiltonians and ans\"atze containing at most two-qubit gates, in order for that path to contribute it needs to satisfy $r_- \geq 2(W-1)$, and hence 
\[\abs{\Phi_{\vec\omega}(\vec{\theta})} \approx \theta^{r_-(\vec\omega)} \leq \theta^{2(W-1)},\] given $\abs{\theta} < 1$, meaning that even if the path actually contributes, it does so to a small amount.

\subsection{The original expected risk in PP}
Even if \Cref{eq:Rq_full} is intractable in practice for PP algorithms, we can compute $E\left[\bbra{0} \bm {L_S(\vec\theta)}\kket{0}\right]_{S \sim \mathcal{Q}_{LS}}$ using the same technique as above. We use again the single qubit Haar measure as example of locally scrambling ensemble. Since $\ketbra 0 = \frac1{2^n}\sum_b Z_b$, for $Z_b$ as in \Cref{eq:Zb}, backpropagating $Z_b$ through $\bm S$ gives rise to $3^{W(Z_b)}$ paths
\begin{equation}
\bbra{Z_b} \bm{S^\dag} = \sum_{P\in \mathbb P_{n, b}} \left(\prod_{i=1}^n \Phi_{P_{b_i}}(\vec\phi_i)\right) P,
\end{equation}
where $\mathbb P_{n, b}:=\{P\in\mathbb P_n: P_{b_i}=I \Leftrightarrow b_i =0\}$ is the set of $3^n$ Pauli strings with weight $W(Z_b)$ that are non trivial exactly where $b_i=1$. In order to compute the expectation value of the sampling analytically, the arguments above can be replicated also in this case, and we find
\begin{equation}
E\left[\bbra{0} \bm {L_S(\vec\theta)}\kket{0}\right]_{S \sim \mathcal{Q}_{LS}} = \frac1{2^n}\sum_{P\in \mathbb P_n} \frac1{3^{W(P)}} \bbra{P} \bm {L_S(\vec\theta)}\kket{P}.
\end{equation}
With this result we interpret \Cref{eq:locHaarpref} as the ``weight $1$'' approximation of the full expected risk. It could be possible to consider higher weight approximations, but the number of Pauli strings that one needs to propagate will grow exponentially.

\section{Experimental results}\label{sec:app:experimental}
\begin{figure}
    \centering
    \includegraphics[width=.99\textwidth]{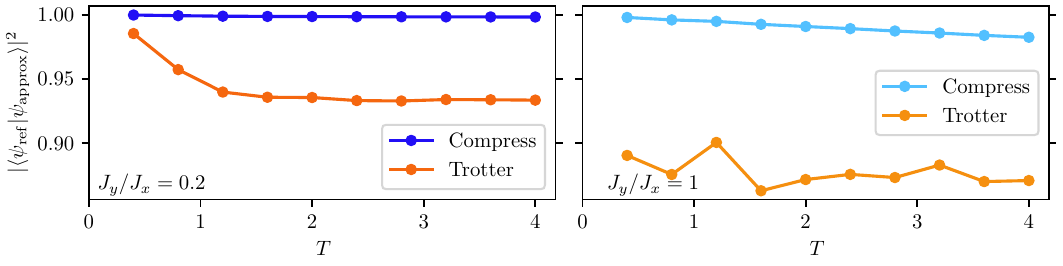}
    \caption{Extended version of \Cref{fig:H1exp}d): fidelities of states prepared using the compressed ansatz or standard Trotterization against a reference state for $J_y/J_x=0.2$ (left) and $J_y/J_x=1$ (right). The three leftmost datapoints correspond to $T=0.4, 0.8,1.2$, which are the same simulation times we used in the quantum device implementation.}
    \label{fig:fidelities}
\end{figure}

In order to simulate the dynamics of \Cref{eq:Hbos} on Quantinuum's H1 chip, we mapped the bosonic Hamiltonian \eqref{eq:Hbos} to a spin $1/2$ XY Hamiltonian. The compression was then performed as discussed in the main text. Using a statevector simulator, we compare the fidelity of a state prepared using either multiple applications of the compressed ansatz or by standard Trotterization against a reference state prepared using a much finer Trotterization. The initial state is chosen to be a state with charge filled only at two sites at the center of the lattice (in $x$ direction) (see \Cref{fig:H1exp}a)). The fidelities are displayed in \Cref{fig:fidelities}. This comparison between the compressed and Trotter fidelities is particularly fair because they have identical depth, and \Cref{fig:fidelities} clearly shows that the compressed ansatz achieves much higher fidelities than the standard Trotterization for identical resources.

For the two experiments $2000$ ($J_y/J_x=1$) and $1500$ ($J_y/J_x=0.2$) sampling shots were allocated, but we post-selected the bistrings enforcing boson number conservation, as it's expected by the physical model and it therefore acts as a mechanism protecting against number-altering errors in the device. For the longer simulation times, we report that over half the shots were discarded. In \Cref{tab:bosons_extra} we give an overview of the comparison between the compressed and the Trotter circuits: for the noisy setup we report the average occupation number error $\overline{|\langle n_j\rangle(T) - \langle n_j\rangle_\mathrm{ref}(T)|}$ and the percentage of ``physical'' (number preserving) samples $p_\mathrm{phys}$ for both the H1 results of the compressed circuit and the results from H1 noisy emulator for the Trotter circuit. For the noiseless simulations with the statevector simulator, we report the infidelites $1-|\braket{\psi_\text{ref}}{\psi_\text{trotter}}|^2$.

\begin{table}
    \centering
    \begin{tabular}{cc|cc|cc|c|c}
    \toprule
    & & \multicolumn{4}{c|}{Noisy} & \multicolumn{2}{c}{Noiseless}\\
    \midrule
    & & \multicolumn{2}{c|}{Experiment} &  \multicolumn{2}{c|}{Emulator} & & \\
    & & \multicolumn{2}{c|}{Compress} &  \multicolumn{2}{c|}{Trotter}& Compress & Trotter\\
    \midrule 
    $J_y/J_x$ & $T$ & $\overline{|\langle n_j\rangle(T) - \langle n_j\rangle_\mathrm{ref}(T)|}$ & $p_\mathrm{phys}$ & $\overline{|\langle n_j\rangle(T) - \langle n_j\rangle_\mathrm{ref}(T)|}$ & $p_\mathrm{phys}$ & $1-|\braket{\psi_\text{ref}}{\psi_\text{compress}}|^2$& $1-|\braket{\psi_\text{ref}}{\psi_\text{trotter}}|^2$\\
    \midrule 
    $0.2$ & $0.4$ & $4.6\cdot 10^{-3}$ & $0.822$ &$5.7\cdot 10^{-2}$ & 0.769 & $2.3\cdot 10^{-4}$ & $1.5\cdot 10^{-2}$\\
    $0.2$ & $0.8$ &$7.0\cdot 10^{-3}$ &$0.632$ & $9.9\cdot 10^{-2}$&0.380 
    & $7.2\cdot 10^{-4}$ &$4.3\cdot 10^{-2}$ \\
    $0.2$ & $1.2$ & $9.7\cdot 10^{-3}$ &$0.483$ &$8.8\cdot 10^{-2}$ &0.173  & $1.1\cdot 10^{-3}$ & $6.0\cdot 10^{-2}$\\
    \midrule 
    $1$ & $0.4$ &$3.9\cdot 10^{-3}$ &$0.818$ & $7.2\cdot 10^{-2}$& $0.646$  & $2.1\cdot 10^{-3}$ & $1.1\cdot 10^{-1}$\\
    $1$ & $0.8$ & $5.7\cdot 10^{-3}$ &$0.613$ & $9.6\cdot 10^{-2}$& 0.341 & $4.0\cdot 10^{-3}$ & $1.2\cdot 10^{-1}$\\
    $1$ & $1.2$ &$6.2\cdot 10^{-3}$ &$0.480$ &$7.3\cdot 10^{-2}$ & 0.147 & $5.2\cdot 10^{-3}$ &$1.0\cdot 10^{-1}$ \\
    \bottomrule 
    \end{tabular}
    \caption{Noisy and noiseless comparison of of states prepared with the compressed and the Trotter circuit. For the noisy case we report the average occupation number error $\overline{|\langle n_j\rangle(T) - \langle n_j\rangle_\mathrm{ref}(T)|}$ and the percentage of number preserving samples $p_\mathrm{phys} = N_\mathrm{number~preserving~shots}/N_\mathrm{shots}$. For the noiseless case we report the infidelities $1-|\braket{\psi_\text{ref}}{\psi_\text{trotter}}|^2$.}
    \label{tab:bosons_extra}
\end{table}

\end{document}